\documentclass[final,5p,times,twocolumn]{elsarticle}

\usepackage{amssymb}
\usepackage{amsmath}
\usepackage{amsthm}

\usepackage{graphicx}
\usepackage{booktabs}
\usepackage{algorithm}
\usepackage{algorithmic}
\usepackage{multirow}

\usepackage[pdftex,colorlinks=true,citecolor=blue,linkcolor=blue,urlcolor=blue]{hyperref}

\begin{document}

\begin{frontmatter}

\title{Pocket RAG: On-Device RAG for First Aid Guidance in Offline Mobile Environments}

\author[ustechlab]{Dong Ho Kang\corref{cor1}}
\ead{donghokang@ustechlab.com}

\author[hanyang]{Hyunjoon Lee}
\ead{amo700@hanyang.ac.kr}

\author[ewha]{Hyeonjeong Cha}
\ead{hyeonjeong.cha@ewha.ac.kr}

\author[utexas]{Minkyu Choi}
\ead{minkyu.choi@utexas.edu}

\author[kookmin]{Sungsoo Lim}
\ead{sungsoo.lim@kookmin.ac.kr}

\cortext[cor1]{Corresponding author}

\affiliation[ustechlab]{organization={Ustechlab Research Institute},
            city={Gwangmyeong},
            country={South Korea}}

\affiliation[hanyang]{organization={Department of Computer Science, Hanyang University},
            city={Seoul},
            country={South Korea}}

\affiliation[ewha]{organization={Department of Business Administration, Ewha Womans University},
            city={Seoul},
            country={South Korea}}

\affiliation[utexas]{organization={Department of Computer Science, The University of Texas at Austin},
            city={Austin},
            state={TX},
            country={USA}}

\affiliation[kookmin]{organization={Department of Computer Science, Kookmin University},
            city={Seoul},
            country={South Korea}}

\begin{abstract}
In disaster scenarios or remote areas, first responders often lose network connectivity when providing first aid. In such situations, server-based AI systems fail to provide critical guidance. To address this issue, we present a lightweight, mobile-based retrieval-augmented generation system for small language models (SLMs) that can run directly on Android devices. Our system integrates a mobile-friendly optimized pipeline featuring Hybrid RAG, selective compression, batched prompt decoding, and quantization caching. Despite the model's small size, our RAG-based system achieves 94.5\% accuracy for physical first aid and 97.0\% for psychological first aid. Additionally, we reduce response time from 14.2s to 3.7s, achieving a nearly 4x speedup. These results prove that our system is practical and can deliver reliable first aid guidance even without internet connectivity.
\end{abstract}


\begin{highlights}
\item A resource-optimized AI system that operates within Android's 2GB memory constraint
\item Real-time inference optimization achieving 3.7s TTFT (nearly 4x speedup)
\item 94.5\% accuracy on physical first aid and 97.0\% on psychological first aid datasets
\item Open-source mobile health system supporting modular updates for models and knowledge bases
\end{highlights}


\end{frontmatter}


\section{Introduction}
\label{sec:introduction}

Mobile health technologies have revolutionized first aid delivery by providing portable access to life-saving guidance when professional medical help is unavailable~\cite{Steinhubl2013_mHealth}. Recent studies increasingly highlight the critical role of mobile technologies in bridging the healthcare access gap in resource-constrained environments~\cite{Kubau2025_Humanitarian}. Furthermore, recent AI-driven systems have demonstrated the capacity to support laypersons and first responders by providing real-time first aid instructions during critical incidents~\cite{abozahhad2025development}. However, a critical vulnerability persists in these advancements: the heavy reliance on cloud-based architectures. This dependency becomes a fundamental limitation in disaster scenarios or remote areas where network connectivity cannot be guaranteed, precisely when first aid guidance is most needed.

\begin{figure}[t]
\centering
\includegraphics[width=\columnwidth]{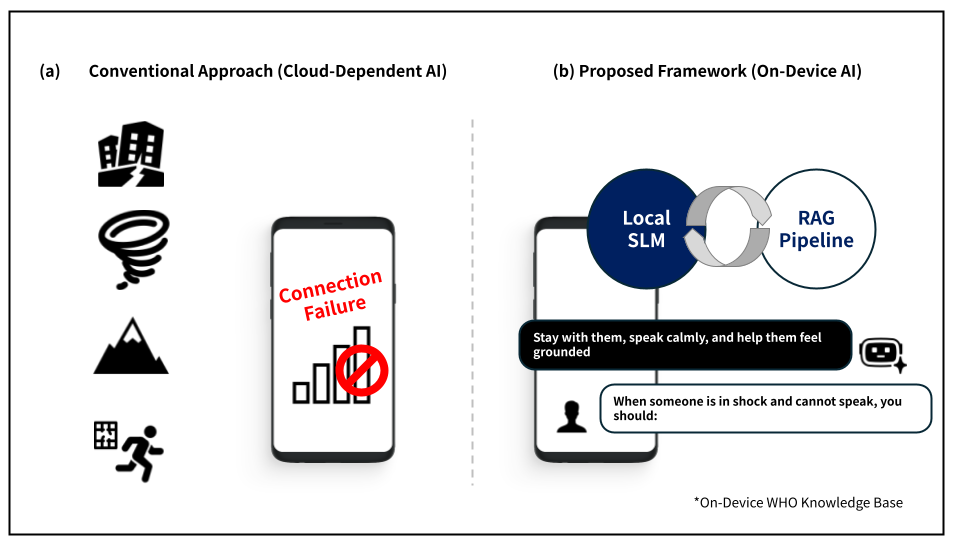}
\caption{Comparison between conventional cloud-dependent AI systems and our proposed on-device framework for first aid scenarios. Traditional approaches (a) fail when connectivity is compromised during disasters---precisely when first aid guidance is most critical. Our framework (b) addresses this limitation by embedding a Local SLM with RAG pipeline directly on mobile devices, enabling reliable access to WHO first aid knowledge without internet dependency.}
\label{fig:motivation}
\end{figure}

Although recent developments in Small Language Model (SLMs) have demonstrated potential for mobile AI~\cite{yang2025qwen3technicalreport,Phi3TechnicalReport2024}, there are two critical barriers to practical application in mobile environments. First, SLMs generally underperform LLMs, which limits their applicability in domains that require high reliability, such as first aid guidance.
Second, mobile devices have extremely limited resources. In particular, the Android system strictly caps memory usage per app (at 2GB ~\cite{Li2018_EdgeComputing}), making it difficult to run resource-intensive models. Also, computational overload on mobile CPUs results in significant latency, which risks rendering the system ineffective during time-sensitive first aid situations. For these reasons, server-based AI or static checklist-based applications are preferred over SLMs.

To overcome these limitations, fine-tuning and retrieval-augmented generation (RAG) are considered to increase the performance of language models. However, for SLMs, related research remains scarce, and most existing services are constrained by their reliance on server-based infrastructure. Consequently, open frameworks for first aid RAG that operate in mobile environments—specifically on the Android platform, which accounts for 70\% of the global market—are virtually non-existent. Therefore, we propose an RAG-based mobile AI system optimized for first aid guidance in offline scenarios.

\paragraph{Our Contribution.}  Our study demonstrates that a local SLM with a resource-optimized RAG framework can deliver trustworthy first aid guidance within practically viable response times. Our contributions are fourfold:

\begin{enumerate}
\item \textbf{Resource-optimized RAG Architecture:} We suggest a mobile-native pipeline that operates strictly within the 2 GB memory ceiling. By implementing a \textit{Hybrid Retrieval Mechanism} (lexical filtering + semantic reranking) and an \textit{Adaptive Memory Manager}, we ensure system stability without compromising retrieval quality.

\item \textbf{Real-time Inference Optimization:} We apply the latency reduction techniques such as Batched Prompt Decoding, KV Cache Quantization, and Selective Context. Together, these optimizations reduce the Time-to-First-Token (TTFT) from 14.2 seconds to \textbf{3.7 seconds}---a nearly \textbf{4x speedup} that brings the system within striking distance of cloud-based responsiveness.

\item \textbf{Physical and psychological first aid support:} Beyond traditional physical first aid, our study evaluates on both Physical and \textbf{Psychological First Aid} datasets derived from WHO guidelines. Our experiments with three SLMs (e.g., Qwen3 0.6B) demonstrate versatile capability in handling diverse first aid scenarios for laypersons and first responders.

\item \textbf{Contribution to AI for Social Good} We release our source code as open source for the open and sustainability mobile health system. Our framework is designed for modular, enabling seamless updates to both the underlying language models and medical knowledge sources easily. It allows that the application can easily change the new medical standards pdf or newly developed SLM with minimal technical barrier.
\end{enumerate}

This paper is organized as follows. Our paper reviews related work in mobile health and on-device AI Section~\ref{sec:related_work}). Section~\ref{sec:methodology} presents our proposed mobile RAG framework, detailing the resource-efficient architecture. Section~\ref{sec:experiments} presents comprehensive results on accuracy and system efficiency. Finally, section~\ref{sec:discussion} discusses the interesting point during our experiments and limitation. Finally, we conclude our papers ~\ref{sec:conclusion}.

\section{Related Work}
\label{sec:related_work}
Our work is multidisciplinary research at the intersection of mobile health applications, SLM, and RAG, motivated by the need for AI under emergency situation. Although prior studies have advanced these individual components, an unified offline solution that satisfies the unique demands of disaster scenarios is in the early stage.

Mobile technology has changed significantly in the expansion of emergency healthcare access~\cite{Steinhubl2013_mHealth}. Early research focused on rule-based Clinical Decision Support Systems (CDSS) or trauma management, but recent efforts have increasingly integrated AI to broaden these capabilities~\cite{Chen2023CDSS}. For instance, Abo-Zahhad et al.~\cite{AboZahhad2025_FirstAidAR} developed an AI-powered AR glasses system for real-time first aid guidance. Tahernejad et al.~\cite{tahernejad2024application} analyzed AI based triage systems for disaster scenarios. These studies demonstrate the potential of AI and mobile devices for emergency situation.

However, a critical vulnerability in these evolutions are the severe dependence on cloud-based architectures. Kubau and Utulu~\cite{kubau2025exploring} highlight that humanitarian and emergency environments are often ``resource-constrained settings'' where network connectivity cannot be guaranteed. Although Xu et al.~\cite{xu2024ondevice} suggested edge-based AI to mitigate this, a comprehensive emergency guidance solution capable of operating independently in completely disconnected environments is still absent.

To address these connectivity constraints, implementing AI directly on the device is essential. However, mobile devices operate under strict limitations regarding computational power and memory. To overcome this, Frantar et al. and others have advanced model quantization techniques~\cite{frantar2023gptqaccurateposttrainingquantization,dettmers2023qloraefficientfinetuningquantized}. Furthermore, recent studies demonstrate that optimized Small Language Models (SLMs) can achieve reasoning capability but provide the small memory footprint~\cite{Wang2024_SLMmHealth}. Liu et al. showed through ``MobileLLM'' that architectural optimizations prioritizing ``depth over width'' allow SLMs to perform effectively even under resource constraints~\cite{Liu2024_MobileLLM}. Our study applies these SLM optimization techniques specifically to the emergency medical domain.

Although SLMs are efficient, their reduced parameters often lead to limited knowledge capacity~\cite{Xu2024_OnDeviceLLMReview} and it causes performance drop. To complement the accuracy, strategies such as fine-tuning or Retrieval-Augmented Generation (RAG) are required. Fine-tuning SLM sounds like tempting. However, data preparation and  retraining for every knowledge update are barrier for non developer users. Instead of finetuning, RAG offers a modular advantage: the system can be updated simply by replacing the external document corpus (e.g., PDF manuals) without modifying model weights. However, existing RAG applications in healthcare do not tested offline constraints. For instance, Moser et al.~\cite{moser2025pipeline} applied RAG to emergency medicine but assumed persistent connectivity. Ren et al.~\cite{Ren2024_RAGMobileEdge} explored mobile edge RAG that still relies on wireless networks. A systematic review by Amugongo et al.~\cite{amugongo2025retrieval} confirms that most healthcare RAG studies depend on cloud resources, making them unsuitable for disconnected environments.

SLM relies heavily on semantic retrieval, the efficiency of the embedding model is as critical for performance. To implement a mobile-native RAG system, the embedding module must be lightweight to fit within the memory budget alongside the SLM. Unlike traditional heavy models, recent lightweight embeddings such as BGE small~\cite{Chen2024_BGEM3} and GTE tiny~\cite{Li2023_GTE} provide high semantic search performance with a minimal memory footprint. We leverage these state-of-the-art lightweight embeddings to optimize the RAG pipeline within strict mobile memory constraints.

To wrap up, existing works have advanced individual elements mobile health, model compression, and RAG but they have not integrated into an unified system tailored for internet denied emergency scenarios. To the best of our knowledge, this is the first study to apply and evaluate an on-device retrieval-augmented small language model specifically optimized for emergency first-aid guidance under mobile resource constraints.

\section{Methodology}
\label{sec:methodology}

Our study presents a resource-constrained, offline-first RAG framework specifically designed for emergency response on standard mobile devices. There are two challenges of strict memory limitations (approx. 2 GB per application on Android) and the need for real-time latency (Time-to-First-Token $< 5$s) without relying on external connectivity.

\begin{figure}[t]
\centering
\includegraphics[width=\columnwidth]{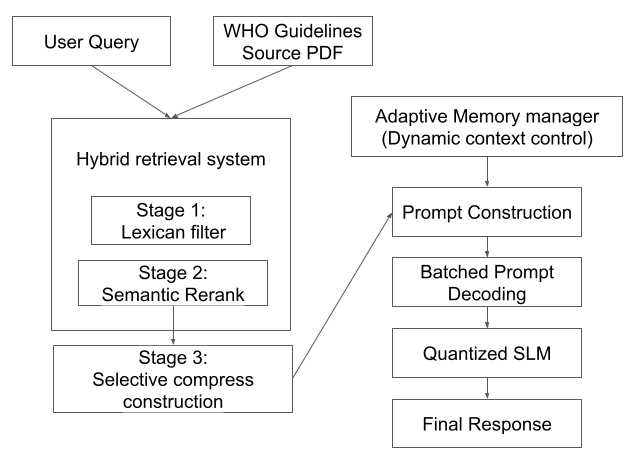}
\caption{System architecture showing the on-device RAG pipeline (left) and evaluation data preparation workflow (right). The enhanced pipeline incorporates \textbf{Selective Context} compression and \textbf{KV Cache Quantization} to maximize responsiveness within the 2 GB memory constraint.}
\label{fig:system_architecture}
\end{figure}

\subsection{System Architecture}

Figure~\ref{fig:system_architecture} illustrates our pipeline, which operates entirely within the mobile application sandbox. The architecture is governed by a strict memory budget constraint:
\begin{equation}
M_{\text{total}} = M_{\text{model}} + M_{\text{index}} + M_{\text{KV}} + M_{\text{runtime}} \leq 2.0\text{ GB}
\end{equation}
where $M_{\text{model}}$ denotes the quantized SLM weights, $M_{\text{index}}$ the vector retrieval index, $M_{\text{KV}}$ the dynamic key-value cache, and $M_{\text{runtime}}$ the application overhead.

\subsection{On-Device RAG Pipeline}

\subsubsection{Document Ingestion and Indexing}

WHO emergency care manuals are converted into retrieval-ready chunks under the Android on-device constraints:
\begin{itemize}
    \item \textbf{Preprocessing:} PDF/HTML is normalized to plain text, de-duplicated, and stripped of boilerplate (headers/footers). Domain terms (e.g., ``airway'', ``hemorrhage'', ``shock'') are preserved as keywords for lexical filtering.
    \item \textbf{Chunking:} Text is segmented into 300 token windows with a 50-token overlap to retain context continuity. Each chunk stores metadata (section title, page id, domain tag).
    \item \textbf{Lexical index:} A lightweight hashmap maps keywords to chunk ids to enable the Stage~1 filter. The index is capped to $\approx$5,000 entries ($<10$ MB).
    \item \textbf{Vector index:} Chunks are embedded offline with a mobile-suitable encoder (e.g., gte-small or bge-small-en-v1.5) and quantized to INT8. Vectors are stored in a flat index totaling $\approx$120 MB for 8,000 chunks, fitting within the 2 GB budget when combined with the SLM weights.
\end{itemize}

\subsection{Lightweight Hybrid Retrieval Mechanism}

Traditional dense retrieval is computationally expensive for mobile CPUs. We devised a two-stage hybrid retrieval mechanism that balances recall with computational efficiency.

\subsubsection{Stage 1: Lexical Pre-filtering}

To minimize the search space for expensive vector computations, we first apply a lexical filter. The system scans document chunks for domain-critical keywords (e.g., \textit{``cardiac arrest''}, \textit{``hemorrhage''}). A lexical score $S_{lex}$ is computed for each chunk $c_i$ given query $q$:
\begin{equation}
S_{lex}(q, c_i) = \frac{|K_q \cap W_{c_i}|}{|K_q|}
\end{equation}
where $K_q$ is the set of medical keywords extracted from the query and $W_{c_i}$ is the set of words in the chunk. This step typically reduces the candidate set from thousands to fewer than 50 chunks with negligible latency.

\subsubsection{Stage 2: Semantic Re-ranking}

For the filtered candidates, we compute semantic similarity using a quantized embedding model optimized for mobile execution. The final hybrid score $U_i$ determines the top-$k$ context chunks:
\begin{equation}
U_i = \alpha \cdot \frac{E(q) \cdot E(c_i)}{\|E(q)\| \|E(c_i)\|} + (1 - \alpha) \cdot S_{lex}(q, c_i)
\end{equation}
We empirically set $\alpha = 0.6$ to prioritize semantic understanding while maintaining lexical precision.
\subsection{Optimized Mobile Inference Engine}

A critical bottleneck in on-device RAG is the ``prefill'' latency---the time required to process the long retrieved context before generating the first token. We introduce two optimizations to accelerate this phase.

\subsubsection{Stage 3: Selective Context Compression}

To further optimize the prefill latency, out framework apply the \textit{Selective Context} mechanism. Instead of feeding the full retrieved chunks to the model, we apply a lightweight relevance filter that scores sentences based on query keyword overlap and medical domain terms. Only the most relevant sentences are retained, reducing the input token count by approximately 20--40\% while preserving a high information density. It works really well for in terms of latency. 

\subsubsection{Batched Prompt Decoding}

Standard mobile inference engines process prompt tokens sequentially, leading to significant latency for long RAG contexts. We implemented a \textbf{Batched Prompt Decoding} strategy that processes the input prompt in parallel blocks of size $B$ (e.g., 512 tokens). This maximizes the utilization of the mobile CPU's SIMD instructions and improves memory bandwidth efficiency. Our implementation reduces the prefill latency $T_{\text{prefill}}$ from:
\begin{equation}
T_{\text{prefill}} \approx \sum_{i=1}^{L} \tau(1) \quad \rightarrow \quad T_{\text{prefill}} \approx \sum_{k=1}^{\lceil L/B \rceil} \tau(B)
\end{equation}
where $L$ is the context length and $\tau(x)$ is the time to process $x$ tokens.

\subsubsection{Quantization KV Cache}

We utilize 8-bit Post-Training Quantization (PTQ) for the SLM weights (Qwen3 0.6B / Gemma3 1B) to fit within the 2GB memory limitation. Furthermore, we applied Eight-Bit KV Cache quantization. By reducing the precision of the Key-Value cache from FP16 to INT8, we virtually double the effective memory bandwidth during the prefill phase without degrading retrieval accuracy.

\subsection{Memory-Aware Adaptive Runtime}

To prevent Out-Of-Memory (OOM) crashes---a common failure mode in mobile AI---we developed a dynamic memory manager that continuously monitors the Android Heap utilization. The system dynamically adjusts the generation parameters based on the current memory pressure ratio $\rho = M_{\text{used}} / M_{\text{max}}$:
\begin{equation}
T_{\text{max}} = 
\begin{cases} 
1024 & \text{if } \rho < 0.70 \text{ (Safe)} \\
768 & \text{if } 0.70 \leq \rho < 0.85 \text{ (Moderate)} \\
256 & \text{if } \rho \geq 0.85 \text{ (Critical)}
\end{cases}
\end{equation}
This adaptive mechanism allows our application to sustain continuous operation even on devices with aggressive background app killing policies.

\section{Experiments}
\label{sec:experiments}

The experimental evaluation was designed to reflect emergency care settings in which
time constraints and decision accuracy are critical. We constructed two domain-specific evaluation datasets derived from official World Health Organization (WHO) training materials. Accuracy is measured as the proportion of correctly selected answers forms. the android prototype app is developed to check the performance in the mobile system. Figure 3 shows the user interface of the prototype .

\begin{figure}[t]
\centering
\includegraphics[width=\columnwidth]{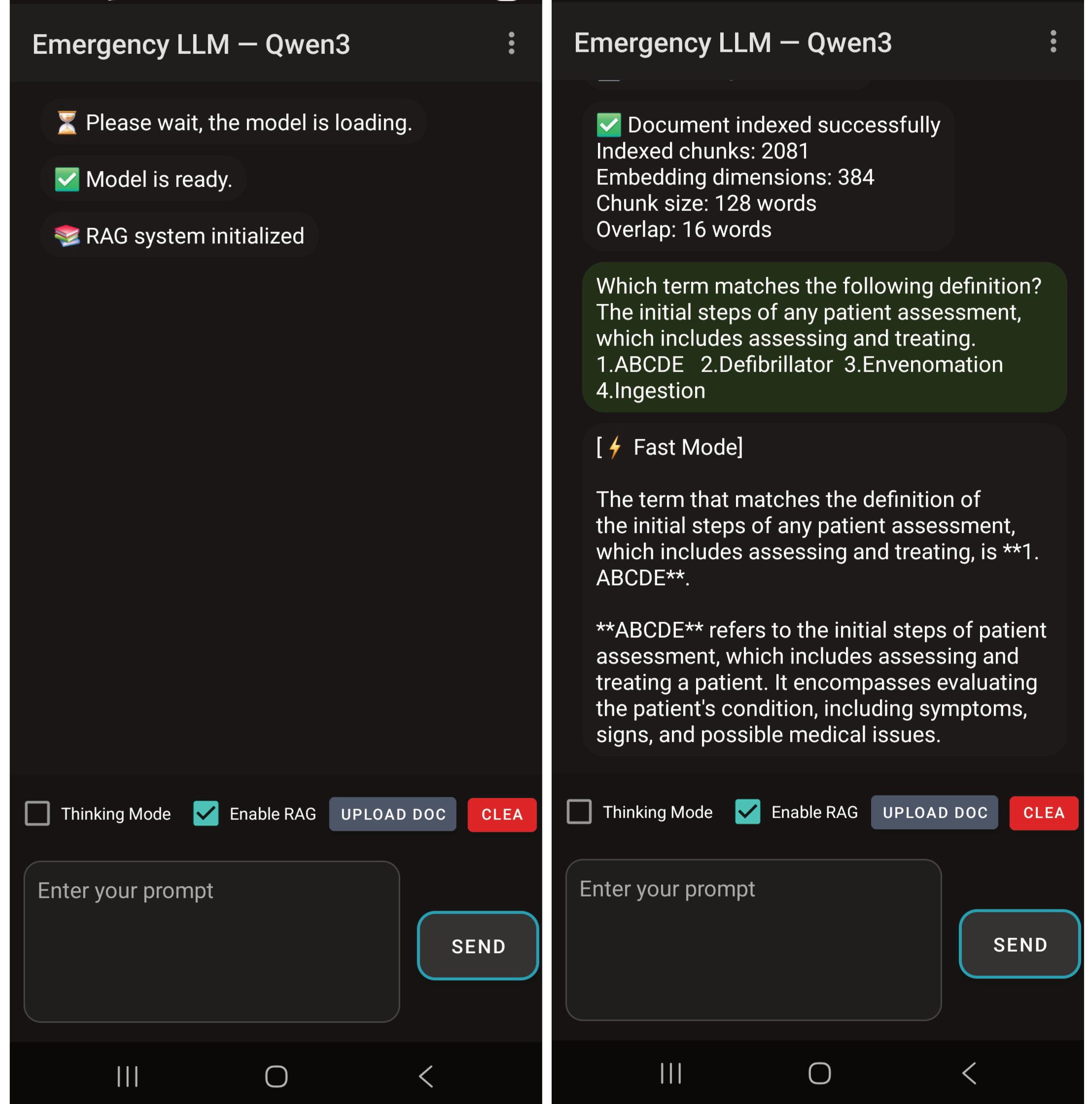}
\caption{Android studio log cat is recorded with speed and resource usage. App is tested the custom made dataset and measure the performance.}
\label{fig:mobile_view}
\end{figure}

\subsection{Pocket Emergency care Dataset}
The BEC dataset was derived from the WHO Basic Emergency Care (BEC) manual, which is commonly used as a reference for frontline emergency response. Relevant procedures and decision rules were converted into 200 multiple-choice questions, each with four candidate answers. The PFA dataset was constructed from WHO training materials on mental health and psychosocial support. It consists of 200 multiple-choice questions addressing key practices, including safety assessment, supportive communication, and referral to available support resources.

\subsection{Accuracy Experimental Setup}

Three small-scale language models were selected for evaluation: Gemma3 1B, Qwen3 0.6B,and Llama3 1B~\cite{llama3_2024,gemma3_2025,qwen3_2025}. Performance differences were analyzed across retrieval configurations and embedding choices, with a focus on how these components affect end-to-end answer accuracy.

\subsubsection{Experiment 1: Physical First Aid (BEC)}

Physical emergency care performance was assessed using questions from the BEC dataset. Each model was evaluated under three inference configurations: a standalone setting without retrieval, retrieval-augmented generation (RAG), and RAG combined with reranking.

\begin{table}[t]
\centering
\caption{Accuracy (\%) on the Physical First Aid dataset (200 WHO-derived multiple-choice questions).}
\label{tab:rag_results}
\begin{tabular}{lrrr}
\toprule
\textbf{Model} & \textbf{Vanilla} & \textbf{RAG} & \textbf{RAG + Rerank} \\
\midrule
Gemma3 1B  & 77.5 & 81.0 & 90.0 \\
Qwen3 0.6B & 80.5 & 88.0 & 90.0 \\
Llama3 1B  & 31.0 & 60.0 & 62.5 \\
\bottomrule
\end{tabular}
\end{table}

Results in Table~\ref{tab:rag_results} show consistent gains from RAG across models. The performance remains strongly correlated with the baseline accuracy of the underlying model, indicating a ceiling on achievable system performance. Although retrieval provides access to task-relevant factual information and reduces incorrect responses, it does not compensate for limitations in the model's core reasoning behavior. For example Llama3 1B was the 31\% accuracy. Our RAG framework boosts its performance from 31 to 62.5. It is almost 31.5 \% accuracy increase. However it is still lower than Gemma3 1B or Qwen3 0.6B which achieve 90.0\% accuracy. 

\begin{table}[t]
\centering
\caption{Accuracy (\%) under RAG + Rerank with different embedding models on the Physical First Aid dataset.}
\label{tab:embedding_results}
\resizebox{\columnwidth}{!}{%
\begin{tabular}{lrrr}
\toprule
\textbf{Embedding Model} & \textbf{Gemma3 1B} & \textbf{Qwen3 0.6B} & \textbf{Llama3 1B} \\
\midrule
all-miniLM        & 90.0 & 90.0 & 62.5 \\
bge-small-en-v1.5 & 86.5 & 90.0 & 62.0 \\
gte-small         & 91.0 & 94.5 & 64.0 \\
\bottomrule
\end{tabular}%
}
\end{table}

Table~\ref{tab:embedding_results} shows that the gte-small embedding consistently yielded the highest accuracy. Qwen3 0.6B combined with gte-small achieved the best performance of 94.5\%.

\subsubsection{Experiment 2: Psychological First Aid}

In the second experiment, we evaluated model performance on the Psychological First Aid dataset.

\begin{table}[t]
\centering
\caption{Accuracy (\%) on the Psychological First Aid dataset (200 WHO-derived multiple-choice questions).}
\label{tab:psy_rag_result}
\begin{tabular}{lrrr}
\toprule
\textbf{Model} & \textbf{Vanilla} & \textbf{RAG} & \textbf{RAG + Rerank} \\
\midrule
Gemma3 1B  & 73.0 & 77.0 & 89.5 \\
Qwen3 0.6B & 89.0 & 94.0 & 97.0 \\
Llama3 1B  & 31.0 & 37.3 & 57.0 \\
\bottomrule
\end{tabular}
\end{table}

Table~\ref{tab:psy_rag_result} shows that Qwen3 0.6B achieved the highest performance with accuracy rising from 89.0\% under Vanilla inference to 97.0\% with RAG + Rerank.

\begin{table}[t]
\centering
\caption{Accuracy (\%) under RAG + Rerank with different embedding models on the Psychological First Aid dataset.}
\label{tab:psy_embedding_results}
\resizebox{\columnwidth}{!}{%
\begin{tabular}{lrrr}
\toprule
\textbf{Embedding Model} & \textbf{Gemma3 1B} & \textbf{Qwen3 0.6B} & \textbf{Llama3 1B} \\
\midrule
all-miniLM        & 89.5 & 95.0 & 58.5 \\
bge-small-en-v1.5 & 89.5 & 97.0 & 57.0 \\
gte-small         & 93.0 & 96.0 & 58.0 \\
\bottomrule
\end{tabular}%
}
\end{table}

Table~\ref{tab:psy_embedding_results} shows that Qwen3 0.6B reached its peak performance with bge-small-en-v1.5 (97.0\%), indicating that embedding--model synergy varies across architectures.

\subsection{General Benchmark Evaluation}

To validate of our lightweight RAG framework beyond domain-specific emergency care datasets, we evaluated our system on standard question-answering benchmarks: SQuAD~\cite{rajpurkar2016squad} and HotpotQA~\cite{yang2018hotpotqa}. These evaluations demonstrate that our resource-optimized approach maintains competitive accuracy while achieving significantly faster response times compared to state-of-the-art mobile RAG systems. About the mobile RAG, there are two algorithm we can compare, EdgeRag~\cite{seemakhupt2024edgerag}and MobileRag~\cite{park2025mobilerag}. 

\begin{table}[t]
\centering
\caption{Comparison with MobileRAG on standard QA benchmarks. Our system achieves competitive or superior accuracy while maintaining a TTFT of 3.7s on mobile devices.}
\label{tab:mobilerag_comparison}
\resizebox{\columnwidth}{!}{%
\begin{tabular}{lrrrrrrr}
\toprule
 & \multicolumn{2}{c}{\textbf{SQuAD}} & \multicolumn{2}{c}{\textbf{HotpotQA}} \\
\cmidrule(lr){2-3} \cmidrule(lr){4-5}
\textbf{Method} & Acc. (\%) & TTFT (s) & Acc. (\%) & TTFT (s) \\
\midrule
Naive-RAG & 55.4 & 6.8 & 27.7 & 13.2 \\
EdgeRAG & 55.4 & 6.8 & 27.7 & 13.4 \\
Advanced RAG & 56.6 & 7.2 & 27.9 & 13.5 \\
MobileRAG & 56.5 & 5.0 & 27.7 & 11.8 \\
\textbf{Ours} & \textbf{57.0} & \textbf{3.7} & \textbf{36.0} & \textbf{3.7} \\
\bottomrule
\end{tabular}%
}
\end{table}

Table~\ref{tab:mobilerag_comparison} presents a comparison with MobileRAG, a recent state-of-the-art mobile RAG system. On the SQuAD dataset, our framework achieves 57.0\% accuracy, comparable to MobileRAG's 56.5\%, while maintaining a TTFT of 3.7 seconds---faster than MobileRAG's reported 5.0 seconds on a Galaxy S24 device. More notably, on HotpotQA, which requires multi-hop reasoning, our system achieves 36.0\% accuracy, representing an 8.3 percentage point improvement over MobileRAG's 27.7\%, while still maintaining the same 3.7-second TTFT. 

These results demonstrate that our lightweight RAG architecture, despite its resource constraints (operating within 2GB memory and optimized for mobile CPUs), does not sacrifice retrieval quality. The combination of Hybrid RAG (lexical pre-filtering + semantic re-ranking) and Selective Context Compression enables efficient retrieval while preserving accuracy. This finding is particularly significant for mobile deployment scenarios where both computational efficiency and answer quality are critical constraints.

\subsection{System Efficiency Analysis}
\label{subsec:system_efficiency}

RAG framework boost the accuracy but the speed of SLM was bottleneck. Based on our real phone evaluation, latency was severe as 14200 MS with Dense RAG. It raises concerns for deployment in time-sensitive emergency response scenarios. This trade-off underscores the need to consider latency constraints when applying small language models in on-device settings. All measurements were conducted on a Samsung Galaxy S23 Ultra (Snapdragon 8 Gen 2, 12 GB RAM). We evaluated the system performance on a Samsung Galaxy S23 Ultra (Snapdragon 8 Gen 2, 12GB RAM).

\begin{table}[t]
\centering
\caption{On-Device Performance Comparison: Baseline vs. Optimized Framework.}
\label{tab:performance_comparison}
\begin{tabular}{lrrr}
\toprule
\textbf{Configuration} & \textbf{TTFT (ms)} & \textbf{TPS} & \textbf{Speedup} \\
\midrule
Baseline (Sequential) & $14200 \pm 250$ & 22.57 & -- \\
With Batching & $4800 \pm 400$ & 26.52 & 3.0x \\
\textbf{Ours (Final)} & $\mathbf{3748} \pm \mathbf{300}$ & \textbf{34.05} & \textbf{3.8x} \\
\bottomrule
\end{tabular}
\end{table}

As shown in Table~\ref{tab:performance_comparison}, our combined optimizations significantly transformed the system's responsiveness. The initial Batched Prompt Decoding reduced prefill latency to 4.8 seconds. The addition of KV Cache Quantization and Selective Context further drove this down to 3.7 seconds a nearly 4 times speedup compared to the sequential baseline.

\section{Discussion}
\label{sec:discussion}

Experimental results across both the Physical and Psychological First Aid domains highlight several important considerations for designing reliable on-device small language model (SLM) systems for emergency response scenarios.

\subsection{Accuracy and Reliability}

Retrieval-augmented generation (RAG) improved accuracy across all evaluated models and domains. Both RAG and RAG combined with reranking reduced hallucinated responses. However, the effectiveness of RAG was strongly influenced by the underlying capability of the base model. For example, Llama3 1B exhibited performance improvement of approximately 30\% with RAG. It is nearly a two times improvement to compare the original Llama3 1B. However, Llama3 1B is still underperform compared to stronger baseline models such as Qwen3 0.6B and Gemma3 1B. These findings indicate that while RAG serves as an effective auxiliary mechanism, the intrinsic quality of the base SLM remains a critical determinant of overall system performance. Consequently, system architectures should be designed to support easy model replacement and upgrades as improved SLMs become available.

\subsection{The Role of Embeddings in Resource-Constrained RAG}

The results further demonstrate that embedding quality plays a central role in determining system performance under resource-constrained RAG settings. Accuracy variations across different embedding models are particularly consequential in emergency care applications, where errors may carry significant real-world risks. Given the rapid emergence of new embedding models and the fact that certain embeddings interact more effectively with specific language models, system designs should explicitly account for embedding–model compatibility and allow for flexible substitution of embedding components. 

\subsection{Latency and Field Readiness}

One of the major challenges encountered in this study was reducing inference latency for the real-world deployment. When we design the initial research, focused primarily on the limited performance capacity of small language models rather than inference speed. Despite using a 2023 flagship smartphone, RAG based SLMs exhibited substantial delays on mobile hardware. There are two issues. First, Thermal throttling on Android devices causes to unstable performance. If device get heated, android system become slow down. Second, issue is TTFT. In a Dense RAG configuration based on LLaMA.cpp, prefill latency exceeded 14 seconds, rendering the system unsuitable for time-sensitive emergency use.

Experiment showed that latency was dominated by time-to-first-token (TTFT). Even if we had 22 TPS of backward, waiting time of first starting give frustration to user. Although throughput increased from approximately 22 TPS to 34 TPS, this improvement had limited perceptual impact. In contrast, reducing TTFT from 14 sec to 3.7 sec significantly improved user experience. It is still slower than cloud-based solutions, this level of performance is acceptable for non-immediate phases of emergency care without internet connection.

\subsection{Limitations and Future Work}

There are several limitations and possibilities in this study. The use of multiple-choice questions simplifies evaluation but does not capture the complexity of open-ended dialogue that arises in real emergency situations. A more realistic assessment would require free-text generation metrics and interactive role-play settings. While the current system reduced inference latency to 3.7 seconds, additional optimization is still needed to enable smoother interaction in practice. Also, voice interface or different emergency guide book based dataset should be examined in future usecase. Moreover, energy consumption was not a primary focus of this work, despite its importance for battery-powered devices operating in disaster zones, and should be examined in future studies.

\section{Conclusion}
\label{sec:conclusion}

This study investigated the use of retrieval-augmented small language models for on-device emergency support. Across two WHO-derived datasets, RAG consistently improved accuracy. Embedding selection also played a substantial role, with accuracy varying across different model–embedding combinations. Also our RAG framework is able to reduce with optimization strategy such as the hybrid RAG, selective context compression, batched prompt decoding and quantization KV cache. As the result, the findings indicate that lightweight, domain-adapted SLMs can achieve deployable performance for emergency care when retrieval pipelines and embeddings are carefully configured. 

However, several limitations—including the use of multiple-choice benchmarks, a limited set of models, and controlled offline evaluation highlight the need for broader experimental settings and human-centered assessments of usability and trust.

In summary, combining SLM with optimized RAG and embedding pipelines allows on-device AI systems to deliver guideline-consistent emergency support under resource constraints.

\section*{Acknowledgments}

The authors thank the anonymous reviewers for their valuable comments and suggestions.


\end{document}